\documentclass[useAMS,usenatbib]{mn2e}

\def\G1915{GRS $1915$+$105$}
\def\X1550{XTE J$1550-564$}
\def\J1655{GRO J$1655-40$}

\def\ax{IGR~J16320$-$4751}
\def\igr1{IGR~J16318$-$4848}
\def\ig1914{IGR~J19140$+$0951}
\usepackage{txfonts}
\usepackage{color}
\definecolor{red}{rgb}{0.7,0,0}
\definecolor{blue}{rgb}{0,0,0.7}
\def\correc#1{{{#1}}}

\def\integ{{\it INTEGRAL}}
\def\xmm{{\it XMM-Newton}}
\def\nh{$N_{\mathrm{H}}$}

\def\ergcms{erg cm$^{-2}$ s$^{-1}$}
\def\chis{{$\chi^2_{\nu}$}}
\usepackage{amssymb}
\usepackage{epsfig}
\usepackage{epsf}
\usepackage{color}
\definecolor{red}{rgb}{0.7,0,0}
\definecolor{blue}{rgb}{0,0,0.7}

\title[\integ~ and \xmm~ observations of the X-ray pulsar
IGR J16320$-$4751]{\integ~ and \xmm~ observations of the X-ray pulsar
IGR J16320$-$4751/AX~J1631.9$-$4752}

\author[J. Rodriguez et al.]{J. Rodriguez$^{1,2,3}$, A. Bodaghee$^{3,4}$, P. Kaaret$^5$, 
J.A. Tomsick$^6$,  E. Kuulkers$^7$, G. Malaguti,$^8$
\newauthor P.-O. Petrucci$^9$, C. Cabanac$^9$, M. Chernyakova$^{3}$, S. Corbel$^{1,2}$, 
S. Deluit$^{10}$, G. Di Cocco$^8$,
\newauthor  K. Ebisawa$^{11}$, A. Goldwurm$^{1,12}$, G. Henri$^9$, F. Lebrun$^{1,12}$, A. Paizis$^{3,13}$, R. Walter$^{3,4}$
\newauthor \& L. Foschini$^8$.\\
$^1$CEA Saclay, DSM/DAPNIA/Service d'Astrophysique,  91191 Gif sur Yvette, France\\
$^2$ Unit\'e mixte de recherche CEA/CNRS/Universit\'e Paris 7 UMR AIM/7158 \\
$^3$ISDC, Chemin d'Ecogia, 16, 1290 Versoix, Switzerland\\
$^4$Observatoire de Gen\`eve,
Chemin des Maillettes 51, 1290 Sauverny, Switzerland\\
$^5$Department for Physics and Astronomy, University of Iowa, Iowa City, IA 52242, USA\\
$^6$Center for Astrophysics and Space Science, University of California at San Diego, MS 0424,
La Jolla, CA 92093 USA.\\
$^7$ISOC, ESA/ESAC, Urb. Villafranca del Castillo, P.O. Box 50727, 28080 Madrid, Spain.\\
$^8$INAF/IASF, Via Gobetti 101, 40129 Bologna, Italy\\
$^9$Laboratoire d'Astrophysique de l'Obesrvatoire de Grenoble, BP 53X, 38041 Grenoble, France\\
$^{10}$Centre d'Etude Spatiale des Rayonnements  9, avenue du Colonel Roche - Boite postale 
4346 31028 Toulouse Cedex 4, France\\
$^{11}$NASA Goddard Space Flight Center, Code 661, Greenbelt, MD 20771, USA\\
$^{12}$Unit\'e mixte de  recherche APC, 11 place  Berthelot 75005 Paris, France\\
$^{13}$INAF/IASF sezione di Milano, Via Bassini 15, 20133 Milano, Italy\\}

\begin{document}

\date{Accepted . Received ; in original form }

\pagerange{\pageref{firstpage}--\pageref{lastpage}} \pubyear{}

\maketitle
\label{firstpage}

\begin{abstract}
We report on observations of the X-ray pulsar IGR~J16320$-$4751 (a.k.a.
AX~J1631.9$-$4752) performed simultaneously with \integ\ and \xmm. We refine the source position and 
identify the most likely infrared counterpart. Our simultaneous coverage 
allows us to confirm the presence of X-ray pulsations at $\sim 1300$ s,
that we detect above 20 keV with \integ\ for the first time. The pulse fraction
is consistent with being constant with energy, which is compatible with
a model of polar accretion by a pulsar. We study the spectral properties 
of \ax\ during two major periods occurring during the simultaneous coverage 
with both satellites, namely a flare and a non-flare period. 
We detect the presence of a  narrow 6.4 keV iron line in both periods. 
 The presence of such a feature is typical of supergiant wind accretors such 
as Vela X-1 or GX 301$-$2. We inspect the 
spectral variations with respect to the pulse phase during the non-flare period, and 
show that the pulse is solely due to variations of the X-ray flux 
emitted by the source and not to variations of the spectral parameters.
 Our results are therefore compatible with the source 
being a pulsar in a High Mass X-ray Binary. We detect a  soft 
excess appearing in the spectra as a blackbody with a temperature of $\sim$0.07 keV. 
We discuss the origin of the X-ray emission in \ax: 
while the hard X-rays are likely the result of Compton emission produced in the close
vicinity of the pulsar, based on energy argument we suggest that the soft excess is
 likely the emission by a collisionally energised cloud in which the compact object 
is embedded.
\end{abstract}


\section{Introduction}
The INTErnational Gamma-Ray Astrophysics Laboratory (\integ) was launched on 
October 17, 2002 (Winkler et al. 2003). Since then, through regular scans of our 
Galaxy and 
guest observers' observations, about 75 new sources\footnote{An updated list of 
all \integ\ sources can be found at 
http://isdc.unige.ch/$\sim$rodrigue/html/igrsources.html} 
have been discovered mainly with the IBIS \correc{telescope (Ubertini et al.
2003), thanks to its low energy camera} ISGRI (Lebrun et al. 2003). 
Because of its energy range (from 15 keV to
 $\sim1$ MeV), its high angular resolution (12 arcmin), good
 positional accuracy (down to $\sim 0.5$ arcmin for bright sources),
and its unprecedented sensitivity between 20 and 200 keV, IBIS/ISGRI has
helped to answer the question of the origin of the hard
 X-ray background in the Galaxy \citep{lebrun04}. These capacities have 
also allowed to discover many peculiar X-ray binaries characterized by a 
huge equivalent absorption column density (\nh), as high as a few times 
$10^{24}$~cm$^{-2}$ in
 \igr1 \cite{matt,walter}, the first source discovered 
by \integ. Due to the high absorption, most of these sources were
not detected during previous soft X-ray scans of the Galaxy (see e.g. Kuulkers 2005
for a review).\\
\indent \ax\ was detected on Feb. 1, 2003, with ISGRI \citep{tomsick} as 
a hard X-ray source. The source was observed 
to vary significantly in the 15$-$40 keV energy range on time scales of $\sim 1000$ s,
and was sometimes detected  above 60 keV \citep{tomsick,luigi}. 
 Inspection of the X-ray
archives revealed that \ax\ is the hard X-ray counterpart to 
AX J1631.9-4752, observed with {\it ASCA} in 1994 and 1997
\citep{sugizaki}.  Analysis
of archival BeppoSAX-WFC data showed that this source was 
persistent for at
least 8 years (in't Zand et al. 2003). Soon after the discovery
of \ax\ by \integ, an \xmm\ Target of Opportunity was triggered.
This allowed us to obtain the most accurate X-ray position to date
\citep{xmmpaper}, which in particular led to the identification 
of two possible infrared counterparts \citep{tomsick,xmmpaper} (hereafter 
source 1 and 2). 
From this analysis, we suggested that \ax\ is
probably a High Mass X-ray Binary (HMXB) hosting a neutron star
\citep{xmmpaper}. This last point has been reinforced since the
discovery of X-ray pulsations from this source in both \xmm\ and
{\it ASCA} observations \citep{sasha} with a pulse period of about 
1300~s. \correc{Aharonian et al. (2005) reporting results of the survey
of the inner regions of the Galaxy at very high energy with HESS, found  
a new source, HESS J1632$-$478, at a position coincident with \ax, but 
the authors suggest that this could be simply a chance coincidence.} Very recently,
Corbet et al. (2005) reported the discovery of strong modulations of the X-ray
flux seen with Swift, that they interpreted as revealing the orbital period of the system.
The period of 8.96 days that they find is compatible with the system containing an 
early type supergiant (Corbet et al. 2005).\\
We report here observations of \ax\ performed simultaneously with
\integ\ and \xmm\ in August 2004.
The sequence of observations is introduced in the following section 
together with technical information concerning the data reduction. Section 3 
of this paper describes the results which are commented and discussed in
the last part of the paper.

\section{Observations and data reduction}
The journal of the observations can be found in Table \ref{tab:log}. 
The \integ\ observation results from an amalgamation of the observation of \igr1\ (PI Kuulkers,
programme \# 0220007)
with that of \ax\ (PI Foschini, programme \# 0220014). \ax, however, remained the on-axis target for 
this observation.
\begin{table*}
\centering
\caption{Journal of the observations analysed in this paper.}
\begin{tabular}{ccccccc}
\hline
Satellite & Revolution &\multicolumn{3}{c}{Date obs.} & Prop. Id & total duration \\
          &    \#      & Start day & Stop day & (MJD-53000)    & & ($\times 10^3$~s) \\
\hline
\hline
\integ & 226 & 2004-Aug-19 & 2004-Aug-20 & 236.57--237.94& 0220014\& 0220007 & $\sim 118$\\
\xmm & 860  & 2004-Aug-19 &  2004-Aug-20 & 236.55--237.14   & 0201700301& $\sim51$\\
\hline
\end{tabular}
\label{tab:log}
\end{table*}

\subsection{\integ\ Data Reduction}
Our \integ\ observation was performed with the so-called  hexagonal 
dithering pattern \cite{courvoisier}, which consists of a sequence of 7
pointings (called science windows, hereafter: SCW) following a hexagonal 
dithering on and around the position
of the source. Being the on-axis target, \ax\ is always in the fully coded field of view
of the IBIS and SPI instruments, where the instrumental response is optimal.
  The \integ\ data were reduced using the Off-line
Scientific Analysis ({\tt OSA}) v 5.0, with specificities for each instrument 
described below.\\
\indent The data from IBIS/ISGRI were first processed until production of
images in the  20--40 and 40--80 keV energy ranges, with the aim of identifying the
most active sources in the field. From the results of this step, we
produced a catalogue of active sources which was given as an input for 
a second run producing images in the 20--60 and 60--200 keV energy ranges. 
In the latter we forced the 
extraction/cleaning of each of the catalogue sources in order to obtain the most reliable
results for \ax\ (see Goldwurm et al. 2003 for a detailed description of
the IBIS software). In order to check for the presence of the 1300~s X-ray pulsations 
in the IBIS range, we also produced light curves with a
time bin of 250~s \correc{(with the {\tt{OSA5.0/ii\_lc\_extract v2.4.3}} module)} 
in the same energy ranges. \\
\indent We produced spectra using two alternative methods. 
The first method is to use the standard spectral extraction from the {\tt OSA}
pipeline. The second method uses the individual images produced in 20 energy bins, to
estimate the source count rate and build the spectrum as explained in 
Rodriguez et al. (2005). This second method can be used to cross check the results 
obtained with the standard spectral extraction. Comparison of the spectra 
obtained with the two methods showed no significant differences, we 
therefore used the spectra obtained with the standard procedure in our spectral 
fits. We used the standard response matrices provided with {\tt OSA 5.0},
i.e. {\tt isgr\_arf\_rsp\_0010.fits} and {\tt isgr\_rmf\_grp\_0016.fits}, the latter 
rebinned to 20 spectral channels.\\
\indent  The source is not spontaneously detected by the JEM-X detector. We tentatively 
forced the extraction of science products at the position of \ax, but a rapid comparison 
with the \xmm\ data showed that the JEM-X products were very likely contaminated by  the 
nearby black hole candidate 4U 1630$-$47 that was in a bright soft state outburst at 
the time of the observation (Tomsick et al. 2005). These data were not considered further.
We did not use the data from the spectrometer SPI, because the 2.5$^\circ$ angular resolution
 did not allow to discriminate the emission of \ax\ from that of e.g. 4U 1630$-$47 (the 
angular separation between the 2 sources is $\sim 40\arcmin$), or  \igr1\ ($\sim 55\arcmin$).
\subsection{\xmm\ Data Reduction}
The \xmm\ data were reduced with the  Science Analysis System
(SAS) v6.1.0. Event lists for EPIC MOS \cite{turner}  and EPIC PN
\cite{struder} cameras were obtained after processing of the ODF data 
with {\tt emchain} and {\tt epchain}.  During the processing, the data
 were screened by
rejecting periods of high background, and by filtering the
bad pixels.  The EPIC MOS were both operating 
in timing mode allowing one to obtain light curves from the central
chip with 1.5 ms resolution. The EPIC PN was operating in full frame
mode, allowing light curves with a time resolution as high as 73.4 ms 
to be obtained. For the PN camera we extracted the spectra and light curves  
from a 35$^{\prime\prime}$ radius  circle centered on the 
source, while background products were extracted from a 
60$^{\prime\prime}$ circular region free of sources (from the same 
chip). \\
\indent Because of the operating mode of the MOS cameras, no
background estimate can be obtained from the central chip where the source 
lies. A ``quick look'' to the lateral MOS chips 
   has shown that the background remained negligible during the whole
   observation.  The latter could
therefore be neglected in the analysis of the MOS data. MOS light curves
were hence extracted from the central chip of both cameras. For the
three EPIC cameras light curves were extracted  in different energy
ranges (2--12 keV, 0.6--2 keV, 2--5 keV, 5--8 keV, 8--12 keV) with the highest
 time resolution achievable. Barycentric correction was applied 
to all light curves. The MOS light curves were then summed using 
the {\tt FTOOLS lcmath}.\\
\indent A redistribution response matrix and an ancillary response file for
the PN spectral products were generated with {\tt rmfgen} and {\tt
  arfgen}. The resulting spectrum was fitted in {\tt XSPEC} v11.3.1 
\cite{arnaud} simultaneously with the spectrum obtained with \integ/ISGRI.

\section{Results}
\subsection{Refining of the X-ray position}
\begin{figure}
\caption{2MASS K$_S$ 1.3$^\prime\times$0.9$^\prime$ image of the field around 
\ax. The 2 circles represent respectively the \xmm\ position given in Rodriguez 
et al. (2003, large circle), and the refined position discussed in the text.
Infrared sources 1 and 2 are the candidate counterparts.}
\epsfig{file=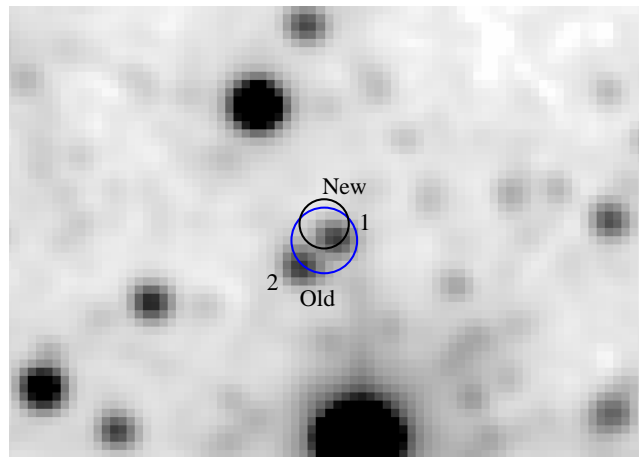,width=\columnwidth}
\label{fig:chart}
\end{figure}

Given the long exposure time (Table \ref{tab:log}) and the high flux from the source
we tried to obtain a better estimate on the position in order 
to possibly discriminate between the 2 candidate counterparts 
\cite{xmmpaper}. For this purpose we used the {\tt edetect\_chain} 
task after having extracted PN images in 5 energy ranges. The latter 
were further rebinned so that an image pixel had a physical size 
of 4.4$^{\prime\prime}$. The best position obtained with this method
is RA$_{J2000}$=248.0077$^\circ$ and DEC$_{J2000}$=-47.8742$^\circ$ 
with a nominal 
uncertainty of $\sim 4^{\prime\prime}$. This is about 1.9$^{\prime\prime}$
from the position reported by Rodriguez et al. (2003). To cross check, 
we re-analysed the 2003 \xmm\ observation with the latest calibration files available, and using the same {\tt edetect\_chain} in 
the same energy ranges. As explained in Rodriguez et al. (2003), only MOS 
data are available, and the data need to be filtered  for soft proton 
flares. The best position we obtained is RA$_{J2000}$=248.0081$^\circ$ 
and DEC$_{J2000}$=$-47.8741^\circ$. We therefore can refine the
 X-ray position to RA$_{J2000}$=16$^h$ 32$^m$ 01$^{s}\!.$9 
DEC$_{J2000}$= -47$^\circ$ 52$^{\prime}$ 27$^{\prime\prime}$
with an uncertainty of $\sim 3^{\prime\prime}$ at 90$\%$ confidence.
This new position suggests more strongly that the source labeled 1 in
Fig. \ref{fig:chart} is the genuine counterpart of \ax, 
since source 2 is now outside of the 90\% error box on the X-ray 
position.\\
We compared the X-ray source positions for high significance sources
(likelihood $\geq$ 200) with infrared sources from the 2MASS catalog.  We
found a match within $0.4\arcsec$ between the XMM-Newton source at 
RA$_{J2000}$=16$^h$ 32$^m$ 12$^{s}\!.$306, 
DEC$_{J2000}$=-47$^\circ$ 44$^{\prime}$ 59$^{\prime\prime}\!.$32 
 and a bright 2MASS source with a K magnitude of 9.3 located at 
RA$_{J2000}$=16$^h$ 32$^m$ 12$^{s}\!.$339, 
DEC$_{J2000}$=-47$^\circ$ 44$^{\prime}$ 59$^{\prime\prime}\!.$45.
We found a second close match between the XMM-Newton source at 
RA$_{J2000}$=16$^h$ 31$^m$ 35$^{s}\!.$642, 
DEC$_{J200}$=-47$^\circ$ 51$^{\prime}$ 27$^{\prime\prime}\!.$42 
 and a bright 2MASS source with  a K magnitude of 9.3 located at 
RA$_{J2000}$=16$^h$ 31$^m$ 35$^{s}\!.$634, 
DEC$_{J2000}$=-47$^\circ$ 51$^{\prime}$ 27$^{\prime\prime}\!.$89. 
Allowing a $1\arcsec$ error circle for the XMM-Newton source and taking
into account the surface density of XMM-Newton sources and 2MASS
sources at appropriate magnitude limits, we find that the chance
probability of occurrence of these X-ray/IR source coincidences is
0.3\%.  This increases our confidence that the astrometry of the X-ray
is  correct and that we have identified the correct IR counterpart
to  IGR J16320$-$4751.

\subsection{Timing Analysis}
\begin{figure*}
\centering
\caption{{\bf Left:} The 2--12 keV \xmm/PN light curve of \ax. {\bf Right
  :} The 20--60 keV \integ/ISGRI light curve of \ax. Note that the two data sets
  have different lengths and the light curves have different time
  resolutions. The time bin of the \xmm\ light curve is 100~s, while that of
  \integ/ISGRI is about 3200 s (length of a SCW). Upper limits are given at 
the 3-$\sigma$ level.}
\epsfig{file=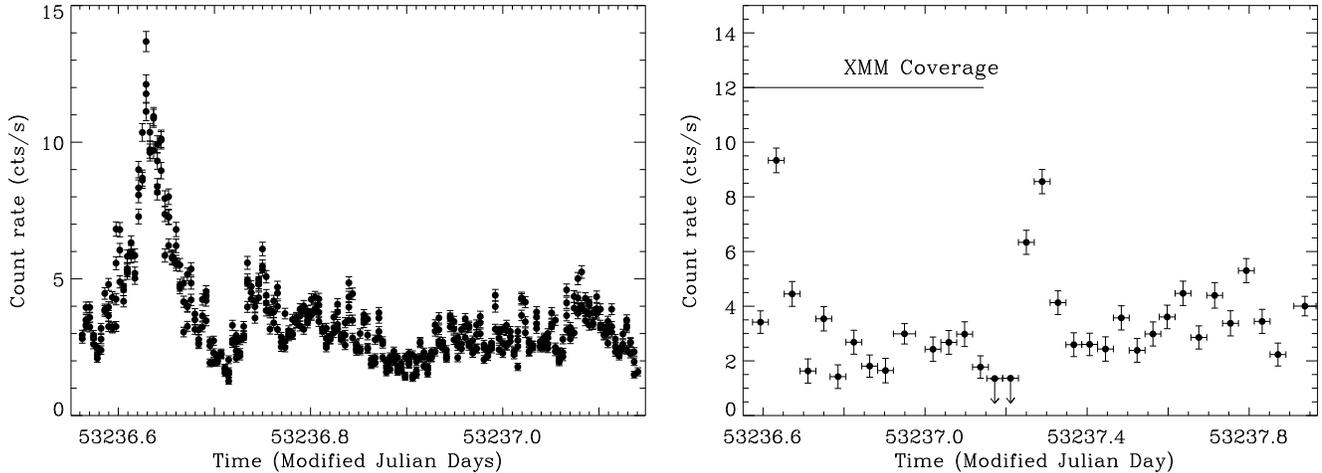,width=\textwidth}
\label{fig:lc}
\end{figure*}

The light curves of \ax\ obtained in different energy ranges with 
 the different instruments are presented in Fig. \ref{fig:lc}. 
As already reported, \ax\ is a variable source on various
time scales, from seconds-minutes (Rodriguez et al. 2003,
Fig. \ref{fig:lc}), to days-months (e.g. Foschini et al. 2004). 
During our observations, the source shows a prominent flare around MJD
53236.6, visible in both light curves (Fig. \ref{fig:lc}). A second
similar flare is visible in the ISGRI 20--60 keV light curve around MJD
53237.3. Unfortunately, our \xmm\ observation does not cover this
period. \\

\begin{figure}
\caption{Lomb-Scargle periodogram of the  2--12 keV PN
  light curve. The insert shows the pulse profile of the 
2--12 keV  light curve obtained by folding the PN light curve at a
period of 1303 s.}
\epsfig{file=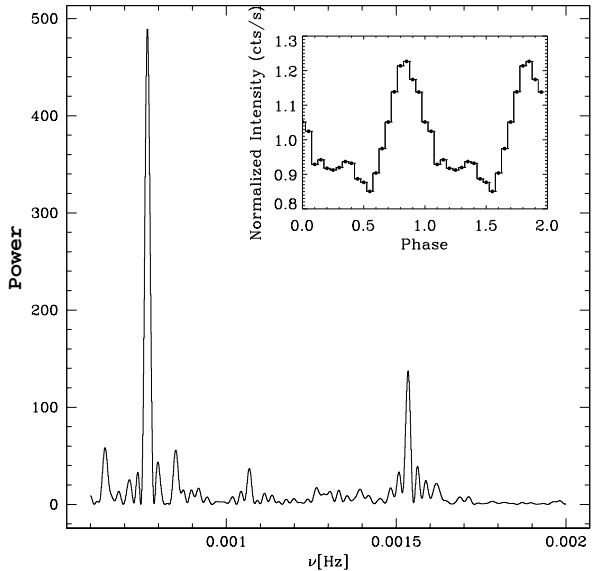,width=\columnwidth}
\label{fig:peak}
\end{figure}

\indent We used the EPIC cameras to study the pulsations of \ax\ already 
reported in Lutovinov et al. (2005). We performed a period search on the 
MOS and PN light curves using the  {\tt  XRONOS} tool 
{\tt efsearch}. Because the source is highly absorbed \cite{xmmpaper}, and
in order to improve the detection of the pulsation we restricted our 
period search to the 2--12 keV range for all EPIC cameras. \correc{The 
Lomb-Scargle of the PN 2--12 keV light curve is reported 
on Fig. \ref{fig:peak}.}

\begin{figure}
\caption{Energy dependence of the pulse fraction, obtained after folding the \xmm\ and \integ\ light
curves. The upper limit at low energy is given at the 3-$\sigma$ level.}
\epsfig{file=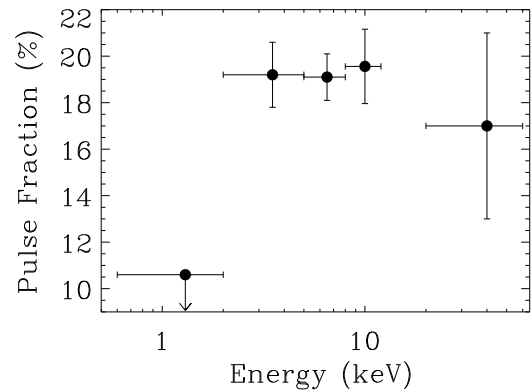,width=\columnwidth}
\label{fig:pulsefrac}
\end{figure}

A very prominent peak is visible with both PN and MOS detectors. 
 Fitting the peak with Gaussian 
profiles lead to best values of 1303.8$\pm$0.9~s (\correc{$7.669 \pm 0.005\times10^{-4}$ Hz}) 
and 1302.0$\pm$1.1~s (\correc{$7.680 \pm 0.006\times10^{-4}$ Hz}) 
for MOS and PN respectively. The errors are calculated from the  periodograms 
using the method developed by Horne \& Baliunas (1986). These values are 
in complete agreement with the 
best value reported by Lutovinov et al. (2005), therefore 
confirming the identification of the pulsation at $\sim$1300~s.\correc{A fainter peak, 
is found at a period half that of the main feature (Fig. \ref{fig:peak}), that identifies it 
as a first harmonic of the pulse period.} 
We then folded the PN light curve with a period of 1303~s (the mean  
of the of the PN and MOS values). \correc{The folded light curve is shown in  
Fig. \ref{fig:peak} (insert)}. 
The pulse fraction (defined as 
$P=\frac{I_{max}-I_{min}}{I_{max}+I_{min}}$, where $I_{max}$ 
and $I_{min}$ represent respectively the intensities at the 
maximum and the minimum of the pulse profile) is $18.11\pm0.7\%$ 
between 2 and 12 keV for the main pulse.
 The pulsation is also visible in the ISGRI 20--60 keV light curve. Folding the light 
curve at a period of 1303~s leads to a pulse fraction between 20 and 60 keV of $17\pm4\%$.
\correc{We  produced  light curves in several energy ranges and folded them with a period of 1303 s,
in order to study the energetic dependence of the pulse fraction. As can be seen on Fig. \ref{fig:pulsefrac} apart from the possible non-detection of pulsations below 2 keV, the pulse amplitude is compatible with
being flat from 2 to 60 keV. The reported upper limit below 2 keV is 10.5\%.}

\subsection{Spectral Analysis}
In order to avoid  mixing different spectral properties together, 
we separated the observation in two different time regions, the first 
corresponding to the first flare (where we isolated $\sim3.6$~ks), 
and the second corresponding to the end of the  simultaneous coverage by 
\xmm\ and \integ, i.e. the last $\sim30$~ks of the \xmm\ observation.
The spectra were fitted  between 0.6 and 12 keV (\xmm), and 20 and 80 keV 
(\integ). No relative normalization constant was included in any of the fits. \\

\subsubsection{Spectral analysis of the non-flaring period} 
\label{sec:averaged}
\begin{figure*}
\caption{\xmm/PN and \integ/ISGRI count spectra {\bf left:} non-flare spectra {\bf right:} flare spectra. In both cases the continuous line represents the best fit model ({\tt edge*wabs*highecut*(powerlaw+gauss)} see the text for the details of the fitting). Note that the same vertical scale is 
used for both.}
\epsfig{file=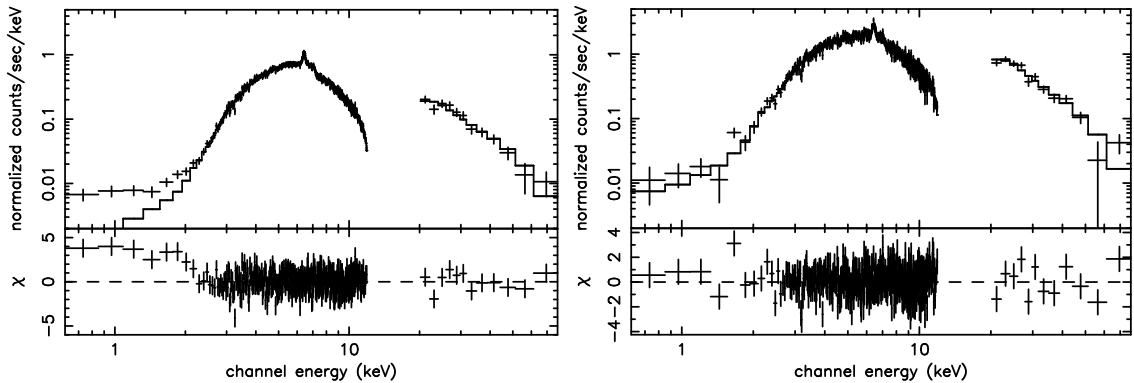,width=15cm}
\label{fig:spec1}
\end{figure*}

Several models were tested in the course of the analysis starting first with 
phenomenological models. A single absorbed power law  leads to a poor reduced $\chi^2$ 
(in the remaining of the paper  \chis\ stands for reduced $\chi^2$). In fact a large 
deviation is seen at high energy, indicative of a high energy cut-off. Replacing the 
single (absorbed) power law by a power law with a high energy cut-off improves the fit,
although the \chis\ is still poor, and high  residuals are found around 6.4 keV. 
 We obtain a relatively acceptable fit with a model consisting 
of an absorbed power law with a high energy cut-off, and a Gaussian emission line 
at $\sim 6.4$ keV. The \chis\ is 1.48 for 441 degrees of freedom (dof). 
This value and  some remaining significant residuals indicate that an improvement is achievable.
This is particularly true  below 2 keV, where a significant excess is detected,  and 
around 7 keV. To account for the latter we included an iron edge in the model.
The fit is significantly improved with \chis=1.20 (for 439 dof). 
 The best fit parameters
are reported in Table \ref{tab:fitparam1} (all errors on the spectral parameters are given at 
the 90\% confidence level). The 2-10 keV unabsorbed flux is 
9.22$\times10^{-11}$ \ergcms, and the 20-100 keV flux is 2.33$\times10^{-10}$ \ergcms.\\
\indent The large  excess still visible below 2 keV (Fig. \ref{fig:spec1}) seems to be reminiscent  
of  several of the so-called highly-absorbed 
sources detected by \integ\ (e.g. IGR J17252$-$3616, Zurita et al. 2005, 
IGR J16393$-$4643, Bodaghee et al. 2005), and has been seen in other X-ray Binaries 
containing pulsars (e.g. Hickox, Narayan \& Kallman 2004). 
In order to try to understand its origin, we fitted the spectra with different types of 
absorption. The first one (\nh$_{ext}$) corresponds to the absorption on the line of sight 
by the Galaxy and was modeled with the {\tt phabs} model in {\tt XSPEC}, while the second (\nh) 
is modeled by photo-electric absorption with variable abundance cross-sections. All the abundances
where frozen to the solar values except that of iron given the presence of an emission line and 
an absorption edge in the simple model presented earlier. The soft excess is modeled by a 
blackbody emission. The spectral model is 
{\tt phabs*(bbody+vphabs*highecut*(powerlaw+gauss))} in the {\tt XSPEC} terminology.
The value of the interstellar absorption (\nh$_{ext}$) was fixed to the value 
obtained from Dickey \& Lockman (1990) in the direction of the source, i.e. 
\nh$_{ext}$=2.1$\times 10^{22}$ cm$^{-2}$ \correc{while the abundances of elements are set
to solar values}. The fit is good with \chis=1.13 (438 dof).
The intrinsic absorption is 11.8$_{-0.4}^{+0.5} \times10^{22}$ cm$^{-2}$, the   
blackbody temperature is  $0.07_{-0.01}^{+0.04}$ keV, 
and the iron abundance is Z$_{Fe}$=1.50$_{-0.15}^{+0.25}$ Z$_\odot$. 
The other parameters are compatible with those returned from the previous fit.
The (extrapolated) 0.01--10 keV unabsorbed flux of the blackbody (soft excess)  is 
3.7 $\times 10^{-10}$ \ergcms, while the 0.01--100 keV unabsorbed flux of the 
power law component is $\sim 4.7\times 10^{-10}$ \ergcms. 

\begin{figure*}
\caption{\xmm/PN and \integ/ISGRI unfolded spectra {\bf left:} non-flare spectra {\bf right:} flare spectra. In both cases the continuous line represents the best fit model ({\tt phabs*(bbody+vphabs*(comptt+gauss))} see the text for the details of the fitting). Individual components are also represented. Note that the same vertical scale is used for both.}
\epsfig{file=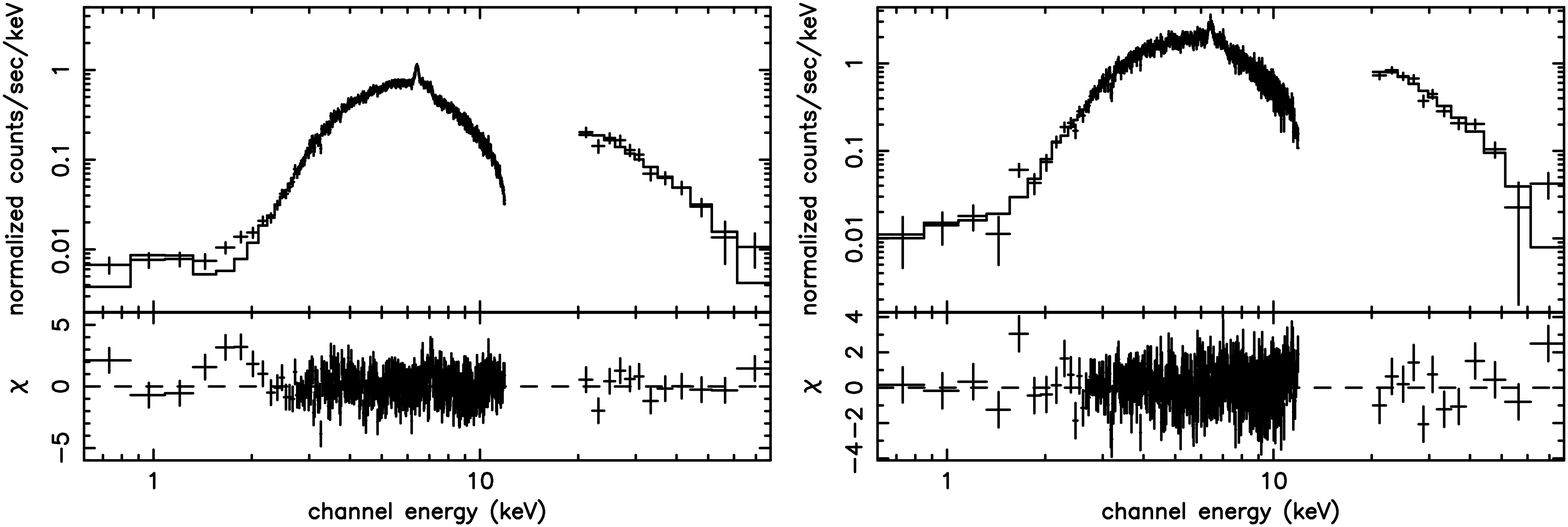,width=15cm}
\label{fig:nuFnu}
\end{figure*}

\begin{table*}
\caption{Best fit parameters obtained from the fit to the 0.6--12 keV and
  20--80 keV \xmm\ and \integ\ flare and non-flare spectra with the simple phenomenological
  model consisting of an absorbed power law with a high energy cut-off
  a  Gaussian and an iron edge. Errors and limits are given at the  90\% confidence level.} 
\begin{tabular}{cccccccccc}
\hline
Spectra & \nh& Cut-off Energy & e-folding & $\Gamma$ &Line (centroid) & Line width & Line eq. width & Edge & Max $\tau$ 
 \\
& ($\times10^{22}$ cm$^{-2}$) & (keV) & (keV) & & (keV) & (keV) & (eV) & (keV) & \\
\hline
\hline
Non-Flare & 15.5$_{-0.6}^{+0.1}$ & 7.1$\pm0.4$ & 13.4$_{-0.9}^{+0.7}$ & $0.24_{-0.12}^{+0.02}$ & $6.409_{-0.005}^{+0.027}$ &$<0.03$& $100_{-14}^{+20}$  &7.21$\pm0.04$& 0.14$\pm0.04$\\ 
\hline
Flare & 11.5$\pm0.5$ & 11.4$_{-0.5}^{+0.9}$ & 10.1$\pm0.7$ & 0.28$_{-0.03}^{+0.07}$ & 6.419$\pm0.014$ & $0.046_{-0.029}^{+0.025}$& $116_{-37}^{+39}$ & 7.3$_{-0.1}^{+0.3}$ & 0.15$_{-0.05}^{+0.02}$ \\
\hline
\end{tabular}
\label{tab:fitparam1}
\end{table*}
\indent Since a cut-off power law is usually interpreted as a signature for thermal 
Comptonisation, we replaced the phenomenological model by a more physical model of 
Comptonisation ({\tt comptt}, Titarchuk 1994). The choice of this model is also
dictated by the fact that it gave a good representation of the \xmm\
spectrum of a previous observation \cite{xmmpaper}, although the lack
of high energy data had prevented us from obtaining a good constraint  on
the temperature of the electrons. A simple absorbed {\tt comptt} model (plus a Gaussian)
represents the data rather well (\chis=1.54 for 442 dof), although the soft excess and an iron edge 
are here again clearly visible. We therefore fitted the data with a model consisting of 
an (externally-) absorbed blackbody plus intrinsically-absorbed Comptonisation (\correc{all abundances 
are set to solar values except the  
iron density of the local (vphabs) absorption that was left as a free parameter}) and a Gaussian line 
({\tt phabs(bbody+vphabs(comptt+gauss))} in XSPEC). Similarly to the previous case,
\nh$_{ext}$ is frozen to the  Galactic value along the line of sight.
We obtain a good representation of the spectrum with \chis=1.15 for 438 dof.
The best fit parameters are reported in Table \ref{tab:complexmodels}, while the 
``$\nu$-$F_\nu$'' spectrum is presented in Fig. \ref{fig:nuFnu}. The blackbody 
(unabsorbed) bolometric flux is 2.2 $\times 10^{-10}$ \ergcms, while the 0.01--100 keV 
unabsorbed \correc{(i.e. external plus intrinsic absorption corrected)} {\tt comptt} 
flux is 4.4 $\times 10^{-10}$ \ergcms.\\

\begin{table*}
\caption{Best fit parameters obtained from the fit to the 0.6--12 keV and
  20--80 keV \xmm\ and \integ\ flare and non-flare spectra with the physical models 
involving thermal Comptonisation. \nh\ is the absorption intrinsic to the source, 
kT$_{bb}$ represents the temperature of the 
soft excess, kT$_{inj}$ the temperature of the injected photon for Comptonisation and  kT$_{e}$
the temperature of the Comptonising cloud.  Errors represent the 90\% confidence level.} 
\begin{tabular}{cccccccccc}
\hline
Spectra & \nh & Z$_{Fe}$ & kT$_{bb}$ & kT$_{inj}$ & kT$_{e}$  & $\tau_\rho$ & Line (centroid) & Line width & Line eq. width\\
& ($\times10^{22}$ cm$^{-2}$) & Z$_\odot$ & (keV) & (keV) & (keV)& & (keV) & (keV) & (eV) \\
\hline
\hline
Non-Flare &   8.9$_{-0.7}^{+0.2}$ & 1.83$_{-0.26}^{+0.30}$  & $0.075_{-0.018}^{+0.024}$ & $1.98_{-0.10}^{+0.15}$  &8.0$_{-0.7}^{+1.0}$& 4.9$_{-0.4}^{+0.5}$ & 6.411$\pm0.006$ & $<0.03$ & $100_{-20}^{+24}$\\ 
       
\hline
Flare & 7.6$_{-1.4}^{+1.2}$ &2.2$_{-0.6}^{+0.9}$ & 0.075 {\emph frozen} & 1.30$_{-0.24}^{+0.31}$ & 6.5$_{-0.3}^{+0.4}$ & 9.8$_{-1.4}^{+1.2}$ & 6.419$_{-0.015}^{+0.014}$ & $0.04\pm0.02$ & $112_{-44}^{+42}$\\

\hline
\end{tabular}
\label{tab:complexmodels}
\end{table*}

We also tried an alternative model involving partial covering by an ionised absorber 
({\tt pcfabs} in {\tt XSPEC}). When leaving the disc temperature free to vary, non-realistic 
values are obtained for its normalization. We then froze $kT_{bb}$ to the best value 
obtained with the other models, i.e. 0.075 keV. The fit is good with \chis= 1.17 for 444 dof. 
The spectral parameters 
are compatible with those obtained with the previous model except that the intrinsic absorption 
is slightly higher here (Table \ref{tab:pcfabs}). Note that the fit indicates that the central 
source is almost completely covered by the absorber.

\begin{table*}
 \centering
\caption{Best fit parameters from the thermal Comptonisation model, modified by a partial-covering 
absorber applied to the 0.6--12 keV and 20--80 keV \xmm\ and \integ\ 
flare and non-flare spectra. Errors and limits denote the 90\% confidence level.  In both cases 
the blackbody temperature of the soft excess is frozen to 0.075 keV.}
\label{tab:pcfabs}
\begin{tabular}{lccccc}  
\hline
Spectra & $N_{\mathrm{H}}$ & kT$_{inj}$ & kT$_{e}$ & $\tau_{\rho}$ & Covering fraction \\
        &  ($\times10^{22}$ cm$^{-2}$)& (keV) & (keV)& & \%\\   
\hline
\hline
Non-flare & 12.2$\pm0.6$ &  1.5$^{+0.05}_{-0.11}$ & 7.0$^{+0.5}_{-0.4}$ & 6.4$\pm$0.7 & $98.4_{-0.7}^{+0.3}$\\  
        \hline	
Flare & 9.2$^{+1.1}_{-1.3}$ & 1.0$\pm0.2$ & 6.6$\pm0.2$ & 9.9$_{-0.7}^{+0.5}$ & $98.0_{-0.5}^{+1.8}$  \\
	      \hline
        \end{tabular}
\end{table*}

\subsubsection{The initial flare}
\label{sec:flare}
\indent We started to fit the spectra with the best (first) phenomenological
model obtained in the previous case, namely an absorbed power law with a high energy
cut-off, a Gaussian line and an iron edge. The \chis\ is 1.06 for 439 dof. The best fit 
parameters are reported in Table \ref{tab:fitparam1}, while the spectra 
are shown in Fig. \ref{fig:spec1} with those of the non-flare period. 
The 2--10 keV unabsorbed flux is 2.39$\times10^{-10}$ \ergcms, and the 20--100 keV 
flux is 4.20$\times10^{-10}$ \ergcms.
It is interesting to note that no soft excess is obvious here, while the value of
\nh\ has significantly decreased. It is possible that the soft 
excess is still present, although  at a level compatible with the
emission from the source (Fig. \ref{fig:spec1}). This could then affect the results of the fits, in 
particular the value of the absorption. In order to compare with the non-flare period,
we fitted the data with the same types of models. We first started with  the 
{\tt phabs*(bbody+vphabs*highecut*(powerlaw+gauss))} model, fixing \nh$_{ext}$ to the Galactic 
value along the line of sight. Given the presence of an iron edge in the simple previous fit, 
we also left the iron abundance free to vary. 
 As one could expect, the parameters of the blackbody are poorly constrained, and the values 
of the parameters are close to those found without the inclusion of the blackbody.
In a second run, we fixed the blackbody temperature to the value found during the non-flare 
period ($0.07$ keV, with the phenomenological model). A good fit is achieved with 
\chis=1.06 (439 dof). The value 
of the intrinsic absorption is still lower than in the non-flare case with 
\nh=$7.8_{-0.8}^{+0.9}\times 10^{22}$ cm$^{-2}$. The cut-off energy is now 
16.2$_{-1.7}^{+1.6}$ keV, while the folding energy is 11.8$_{-1.1}^{+1.2}$ keV. The iron
abundance is quite higher (although compatible within the errors) than in the non-flare 
case with Z$_{Fe}$=1.97$_{-0.52}^{+0.64}$. The other parameters are compatible with those 
obtained with the simple model.
We also allowed the temperature of the blackbody to vary while fixing the value of 
the intrinsic absorption to that found during the non-flare period 
(11.8$\times 10^{22}$ cm$^{-2}$). In this case, however, non credible parameters 
are obtained for the blackbody.\\
\indent Replacing the phenomenological model by a {\tt comptt} also leads to 
a good representation of the spectra with \chis=1.07 for 438 dof. In a first 
fit the blackbody parameters are left free to vary. We obtain an upper limit on the temperature 
of 0.11 keV (90\% confidence), but the normalization is poorly constrained. 
We therefore froze the temperature to the best value obtained during the non-flare 
period (0.075 keV) and refitted the spectra. The \chis\ is 1.07 for 439 dof.
The best fit parameters are almost identical to those obtained
when kT$_{bb}$ is free to vary, and are reported in Table \ref{tab:complexmodels}. Contrary 
to the phenomenological case, the \nh\ value obtained,  is marginally compatible with the 
value obtained from the fit to the non flare period.\\
\indent As in the non-flare case, we replaced the previous model by a model 
involving partial covering. Here again, to fit the soft excess well, a blackbody 
component is required. Since no constraints are obtained on the 
blackbody parameters if the blackbody temperature is left free to vary, we froze the latter 
to 0.075 keV. The fit is good with \chis=1.11 for 442 dof, and the best fit parameters 
are reported in Table \ref{tab:pcfabs}.

\subsection{Phase Resolved Spectroscopy}
\begin{figure}
\caption{Phase diagram of the non-flare period. Only one cycle is represented. The horizontal lines
represent the limit above (respectively below) which the high phase (resp. low phase) PN spectrum has
been extracted.}
\epsfig{file=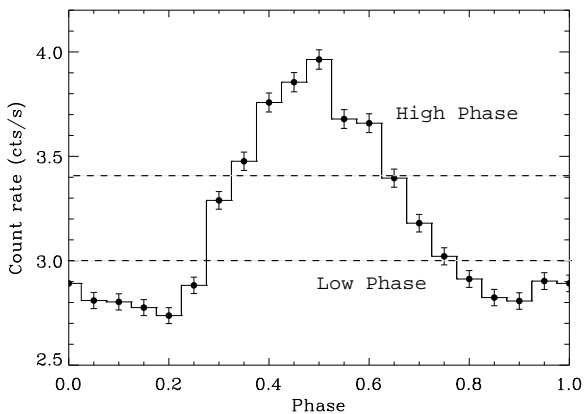,width=\columnwidth}
\label{fig:phaseextract}
\end{figure}

To study the dependence of the spectrum along the phase, we extracted 
PN spectra from the non-flare period in the low part of the 
phase diagram (hereafter low phase) and from the high part of the phase
diagram (hereafter high phase) as illustrated in Fig. \ref{fig:phaseextract},
i.e. the high phase corresponds  to the time when the count rate is higher than 
3.4 cts/s, while the low phase corresponds to the time when the count rate is 
lower than 3 cts/s. \correc{The origin of the phase is taken here as the start of the non-flare period,
which we consider as  the start of the good time intervals for the phase resolved spectroscopy.
The shift of the phase diagram of Fig.~\ref{fig:phaseextract} as compared 
to that of Fig.~\ref{fig:peak} is simply due to the fact that the origin of the 
phase in the former is equal to the beginning of the entire observation. 
Also in Fig.~\ref{fig:phaseextract}, the vertical axis is ``real'' mean counts per second rather 
than normalized ones as it is the case in  Fig.~\ref{fig:peak}}.\\
\indent We focus on the 2--12 keV energy range, since the 0.6--2 keV does not show the presence of 
the pulsation with a rather constraining 3-$\sigma$ upper limit of 
10.5 \%. We fitted the spectra directly with the  
{\tt phabs(vphabs(comptt+gauss))} model. The Galactic absorption (\nh$_{ext}$)
was again fixed to $2.1 \times 10^{22}$ cm$^{-2}$. Since we only consider the \xmm\ spectra here,
no constraint can be obtained for the temperature of the Comptonising electrons. We therefore 
fixed $kT_{e}$ to the value obtained during the fit to the entire non-flare spectra, i.e. 7.8 keV.
In both cases the fits are 
good with \chis=1.18  and \chis=1.01 (422 dof) for the low and high phases, respectively.
The best fit parameters are reported in Table \ref{tab:phaseres}.

\begin{table*}
\caption{Best fit parameters obtained from the fit to the 2--12 keV high and
low phase PN spectra, with the {\tt phabs(vphabs(comptt+gauss))} model.
Errors are given at the 90\% level.} 
\begin{tabular}{cccccccc}
\hline
Spectra & \nh & Z$_{Fe}$  & kT$_{inj}$ & $\tau_\rho$ & Fe Centroid & Eq. width& 2-10 unabs. Flux \\
& ($\times10^{22}$ cm$^{-2}$) & Z$_\odot$ & (keV) & (keV)& keV & eV & ($\times 10^{-10}$ \ergcms)\\
\hline
\hline
Low phase &   9.2$_{-0.8}^{+0.4}$ & 1.7$\pm0.3$ & $1.93_{-0.11}^{+0.17}$  & 5.1$_{-0.6}^{+0.4}$ & 6.411$\pm0.006$ & 104$_{-17}^{+20}$ & 0.86\\ 
       
\hline
High phase & 9.4$\pm1.4$& 1.8$_{-0.5}^{+0.3}$  & $1.9\pm 0.3$  & 4.8$_{-1.5}^{+1.0}$ & 6.44$\pm0.02$ & 72$_{-31}^{+38}$ & 1.04\\

\hline
\end{tabular}
\label{tab:phaseres}
\end{table*}
It is obvious from Table \ref{tab:phaseres} that the spectral parameters of both the 
high and low phase are completely compatible. The only change is clearly related to 
a change of the source flux. \correc{Although the errors on the line equivalent width 
may indicate this parameter does not evolve with the pulse, it is interesting to note 
the  possible decrease of this parameter by $\sim30\%$ between the low 
and high phase. This may suggest that the iron is unpulsed.}

\section{Discussion}
\begin{figure}
\caption{Error contour for the optical depth $\tau$ vs. the electron temperature kT$_e$
from the fit with the {\tt comptt} model and the absorption with variable abundances. The 
crosses mark the location of the best values, while the  contour are the 68, 90 and 99\% 
confidence region.}
\epsfig{file=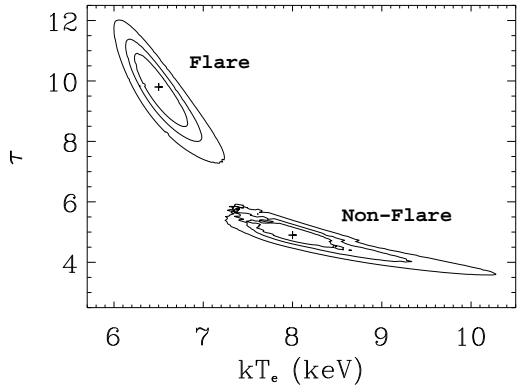,width=\columnwidth}
\label{fig:contour}
\end{figure}

We report here the analysis of simultaneous \xmm\ and \integ\ observations of
the enigmatic source \ax. We focus on the time of strict simultaneous coverage 
by both satellites. We detect  very significant X-ray pulsations at a period of around
1300~s confirming previous findings \cite{sasha}. The pulsation is seen in the \integ/ISGRI light 
curve \correc{above 20 keV}. Apart from the non-detection of the pulsation below 2 keV, no 
particular dependence of the pulse amplitude with the energy is seen.
When studying the phase dependent \xmm\ spectra of the source (in the non-flare)
period, we observe no particular spectral differences between the high-phase and low-phase
spectra. In particular, the intrinsic absorption, temperature of seed photons 
for Comptonisation, and plasma optical depth  remain relatively constant. 
This is compatible with a model of polar accretion by a pulsar. The modulation  of 
the X-ray flux is due to the misalignment of the pulsar spin axis and the pulsar 
magnetic axis. When the pulsar magnetic axis points towards us, the X-ray flux we detect
is enhanced. The pulse period would then be the spin period as already suggested by
Lutovinov et al. (2005). The weak amplitude of the pulse (as compared to other absorbed 
source as e.g. IGR J16393$-$4643, Bodaghee et al. 2005) may indicate that the spin axis 
and the magnetic axis are not highly misaligned, or simply that the angle of the pulsar 
equator with the line of sight is quite low. \correc{The constancy of the pulse amplitude above
20 keV indicates that the Comptonised component is also pulsed. This shows that Comptonisation 
occurs very close to the site of production of the soft photons.}\\
\indent Using two independent \xmm\ observations, we could refine the X-ray position to the
source. This allowed us to further suggest that the infrared source labeled \#1 in Rodriguez et al. 
(2003), and in Fig. \ref{fig:chart} here was the most likely counterpart to \ax.
As discussed in Rodriguez et al. (2003), the magnitudes of this object could be indicative 
of some infrared excess possibly due to circumstellar matter, either a hot plasma or some 
dust.\\
\indent We provide for the first time a  spectral analysis of \ax\ up to 80 keV, with 
significant detection by \integ/ISGRI. The hard X-ray spectrum of the source is 
indicative of a high energy cut-off, whose parameters  can not be constrained when
focusing on the soft X-rays only (Rodriguez et al. 2003). The main emission mechanism
in the source seems to be Compton up-scattering of soft X-ray photons by a hotter 
plasma. The spectral parameters we obtained either with phenomenological fits (Table 
\ref{tab:fitparam1}) or more physical fits to the data are similar to those obtained for
several High Mass X-ray Binaries (HMXB)  (e.g. 4U 1700$-$37, 
Boroson et al. 2003). \correc{If we assume the source is associated with the Norma arm
located between 5 and 10 kpc from the Sun, with a tangent at 8 kpc, we can estimate its 
bolometric luminosity. During the non-flare period, the extrapolated bolometric flux 
of the source is $4.12 \times 10^{-10}$ \ergcms. This leads to a luminosity of 
$3.15\times 10^{36}$ erg/s at 8 kpc, and $1.23\times 10^{36}$ (resp. $4.92\times 10^{36}$) 
for a distance of 5 (resp. 10) kpc. We note that this estimate of the luminosity is compatible 
with the typical ionising luminosity of accretion driven X-ray ($1.2\times 10^{36}$ erg/s) 
pulsars suggested by Bildsten et al. (1997).}  \\
\indent Our spectral analysis allowed us to  
reveal the presence of a soft excess that may be indicative of an additional
medium.  This soft excess is well fitted with a blackbody which temperature is quite low 
and found around 0.07 keV. Such a soft excess seems to be commonly observed in 
X-ray binaries \cite{hickox}, and some other highly absorbed \integ\ sources 
\cite{juan,arash,roland}. This soft excess can have different origins from one system 
to another \cite{hickox}. It could be for example the signature 
of thermal reprocessing of  the harder X-rays produced 
by the accretion of material onto the pulsar either by the inner boundary of an 
optically thick accretion disc, or by an optically thin diffuse cloud. Whatever is the 
exact origin of the medium responsible for  the soft excess, the fact that the 
temperature of the soft excess is different than that of the seed photon for 
Comptonisation rules out a model where this medium would be contained in the Comptonising cloud.
It seems difficult to consider that this medium could be the Comptonising medium giving
birth to the hard X-ray emission. \correc{Given that the Comptonised component shows the pulsation with an 
amplitude similar to that of the soft X-rays}, it is very likely that Comptonisation occurs in the close
vicinity of the pulsar, the seed photons being then probably emitted by the hot 
surface of the compact object, and the up-scattered pulsed Comptonised component emitted by the 
channeled accretion flow.\\
\indent Given the high quality of our data and the long exposure time, we detect 
for the first time a narrow iron line in \ax. This kind of feature is again reminiscent in many 
HMXB, and is also detected in most of the highly-absorbed sources unveiled by \integ, the 
extreme case being IGR J16318$-$4848 \cite{matt,walter,walter2,erik}.
 \correc{The  presence of a narrow fluorescence line, as in e.g. Vela X-1, GX 301$-$2, 4U 1700$-$37, 
or even Cen X-3 (e.g Boroson et al. 2003;  Wojdowski et al. 2003), 
is usually interpreted as fluorescence of iron in a wind or circumstellar matter. This indicates that 
 accretion occurs (at least partly) through a wind. This points towards a high mass companion in \ax\
rather than a low mass one}. The value of the centroid 
obtained during the fit to the non-flare spectra 
corresponds to Fe XIII (House 1969).  We can estimate the distance $R$ between the irradiating
source and the inner radius of the fluorescent shell using $\xi = L/(\rho\times R^2)$, with $L$ 
the luminosity of the source, $\rho$ the density of the gas, and $\xi$ the ionisation
parameter. From Fig. 5 of Kallman et al. 
(2004) we can estimate $log (\xi)<2$ therefore $\xi< 100$. Knowing that $\rho\times R=$\nh, we
obtain $$R> \frac{L}{N_{\mathrm{H}}\times\xi}=10^{11}~\mathrm{cm}\sim 0.07~\mathrm{AU}$$ 
\correc{using the value of the luminosity found assuming a distance of 5 kpc i.e. 
$1.23\times 10^{36}$ erg/s, and $R>0.18$ AU (resp $0.28$ AU) for a distance of 8 (resp. 10) kpc}. 
The inner edge of
the fluorescent shell seems therefore quite far from the inner accretion flow which is again 
compatible with the presence of an additional medium responsible for the soft excess. 
It is interesting to note that when leaving the iron abundance free to vary, it tends 
to values greater than the solar abundance (using the values of the abundance 
of Anders \& Grevesse 1989). 
This could be indicative of the iron origin and the cloud itself
since both seem linked. Iron could have been produced by the evolution of the proto pulsar
in \ax.\\
Given the fact that in all the energy ranges the pulse fraction is roughly
constant, it is very probable that the non-detection of the pulsation below 2 keV indicates that 
the X-ray flux below 2 keV, and consequently the soft excess, do not pulsate. As discussed in 
\cite{hickox}, this clearly rules out the accretion column as the origin of the 
soft excess. This property would suggest that the soft excess is either the emission 
by a collisionally energised cloud or simply the reprocessing of the hard X-rays by a 
diffuse cloud.  In the latter case it seems quite natural to think that 
the absorbing material is the cloud itself.  Because of  energy conservation, 
$F_{bb}\lesssim \Delta F_{comptt}$ where $F_{bb}$ is the unabsorbed bolometric 
flux of the blackbody and $\Delta F_{comptt}$ the difference of the unabsorbed  
to the absorbed ($\sim$bolometric) flux of the Comptonised component. In the non-flare 
case (when the blackbody components are well constrained), our fits lead to 
$\Delta F_{comptt}\sim0.4 \times 10^{10}$ \ergcms.
The unabsorbed bolometric flux of the blackbody component is 2.2 $\times 10^{-10}$ \ergcms.
Hence, the origin of the black body emission cannot be due to reprocessing of the hard X-rays. 
The soft excess would rather be the signature of a collisionally energised 
cloud. \\
\indent We also studied the spectral properties of the source during a flare period 
and compared it to the non-flare period. Although a strict comparison is rendered 
delicate by the presence of the soft excess which is  very poorly constrained during the flare,
it seems that the change during the flare is accompanied by a slight decrease of 
the absorption column density, while  the injected temperature of the photons for 
Comptonisation and that of the Comptonising electrons decrease significantly. Given 
the degeneracy of those two  parameters, the contour plots of $\tau$ vs kT$_e$ are reported 
in Fig. \ref{fig:contour} for both periods. This plot shows that the evolution of the 
Comptonising plasma optical depth and temperature is genuine.
 The ratio of the  $y$($\propto kT_e\tau^2$, \correc{e.g. Titarchuk 1994})  parameters between 
the flare and non flare spectra is $y_{noflare}/y_{flare}$=0.30, indicating a much more 
efficient Compton up-scattering in the case of the flare period.  The short 
time scale on which the flare occurs is more compatible with free-falling phenomena, rather than 
phenomena occuring in an accretion disc where times are of the order of the viscous time scale.
This argument again points to radial accretion in \ax, therefore further suggesting that the system 
is a HMXB. \\
\indent We note that in order to test other possibilities and explain the soft excess, a
model involving partial covering was tested in the course of the analysis. However 
although it can lead to good results,  it has to be noted 
that this model requires the addition of the blackbody to account for the soft excess. The result
of the fits indicate that the covering of the central source of X-ray is quite high and close 
in both cases to its maximum value. The high values of the covering fraction  would tend to 
confirm a picture in which the X-ray source is embedded in a dense material responsible for 
the soft X-ray. The fact that this covering fraction undergoes few variations between the flare and 
the non-flare periods seem to suggest that the flare is not caused by a huge geometrical 
change of this medium. This does not exclude that the flare is
powered by inhomogeneities in the density of the stellar wind close to the
neutron star.
As the size of the region responsible for the absorption and fluorescence
is of the order of the orbital radius, this is where such inhomogeneities in
the wind are in fact expected. The evolution of the absorbing column density
in the model involving partial covering would tend to be in agreement with
a picture, where variations of the local wind density are responsible for
both the variations of the column density and of the hard X-ray flux.
Clearly more observations on this and other highly absorbed sources will shed 
light on the origin of the soft excess in accreting X-ray pulsars. \\ 

\section*{Acknowledgments}
J.R. is specially grateful to J. Zurita for his great help with the timing analysis.
JR acknowledges G. B\'elanger, M. Cadolle-Bel, T. Courvoisier, A. Lutovinov,  
G. Palumbo, M. Revnivtsev for useful discussions on various aspects of the work 
and analysis presented here.  
The authors  thank the anonymous referee for his/her very useful comments.
This work is based on observations with INTEGRAL, an ESA mission with instruments
and science data centre funded by ESA member states (especially the PI
countries: Denmark, France, Germany, Italy, Switzerland, Spain), Czech
Republic and Poland, and with the participation of Russia and the USA,
and observations obtained with XMM-Newton, an ESA science mission with 
instruments and contributions directly funded by ESA Member States and NASA

\label{lastpage}

\end{document}